\documentstyle[aps,multicol,epsf]{revtex}

\begin{document}

\draft

\title{Threading dislocation lines in two-sided flux array decorations}

\author{M.-Carmen Miguel and Mehran Kardar}

\address{Physics Department, Massachusetts Institute of Technology\\ Cambridge,
Massachusetts 02139, USA}

\date{\today}

\maketitle

\begin{abstract}

Two-sided flux decoration experiments indicate that threading dislocation
lines (TDLs), which cross the entire film, are sometimes trapped in metastable 
states.
We calculate the elastic energy associated with the meanderings of a TDL.
The TDL behaves as an {\em anisotropic and dispersive} string with thermal 
fluctuations largely along its Burger's vector. These fluctuations also modify
the structure factor of the vortex solid. Both effects can in principle be used 
to estimate the elastic moduli of the material.

\end{abstract}


\begin{multicols}{2}

Recent two-sided flux decoration experiments have proven an effective technique
to visualize and correlate the positions of individual flux lines on the two
sides of $Bi_2Sr_2CaCu_2O_8$ (BSCCO) thin superconductor films 
\cite{Yao94,Yoon95}. 
This material belongs to the class of high-$T_c$ superconductors (HTSC), whose
novel properties have aroused considerable attention in the last few years
\cite{Blatter}. Due to disorder and thermal fluctuations, the lattice of rigid
lines, representing the ideal behavior of the vortices in {\em clean}
conventional type II superconductors, is distorted.  Flux decoration allows to
quantify the wandering of the lines as they pass through the sample.  
The  resulting decoration patterns also include different topological defects,
such as grain boundaries and dislocations, which in most cases thread the entire 
film.

Decoration experiments are typically carried out by cooling the sample in a 
small magnetic field. In this process, vortices rearrange themselves from a 
liquid-like state at high temperatures, to an increasingly ordered structure, 
until they freeze at a characteristic temperature \cite{Grier,Marchevsky}. Thus, 
the observed patterns do 
not represent equilibrium configurations of lines at the low temperature where
decoration is performed, but metastable configurations formed at this higher 
{\em freezing temperature}. The ordering process upon reducing temperature
requires the removal of various topological defects from the liquid state:
Dislocation loops in the bulk of the sample can shrink, while threading 
dislocation lines (TDLs) that cross the film may annihilate in pairs, or glide 
to the edges. However, the decoration images still show TDLs in the lattice of 
flux lines.
The concentration of defects is actually quite low at the highest applied
magnetic fields $\mbox{\boldmath $H$}$ (around 25$G$), but increases
as  $\mbox{\boldmath $H$}$ is lowered (i.e. at smaller vortex densities).
Given the high energy cost of such defects, it is most likely that they are 
metastable remnants of the liquid state. (Metastable TDLs are also formed 
during the growth of some solid films\cite{TDLs}.)

Generally, a good correspondence in the position of individual
vortices and topological defects is observed as they cross the sample.
Nevertheless, differences at the scale of a few lattice constants
occur, which indicate the wandering of the lines. 
Motivated by these observations, we calculate the extra energy cost
associated with the deviations of a TDL from a straight line conformation.
The {\em meandering} TDL behaves like an elastic string with a
dispersive line tension which depends logarithmically on the wavevector of 
the distortion.
\begin{figure} 
\epsfxsize=7.5truecm 
\vspace*{-0.2truecm} 
\centerline{
\epsfbox{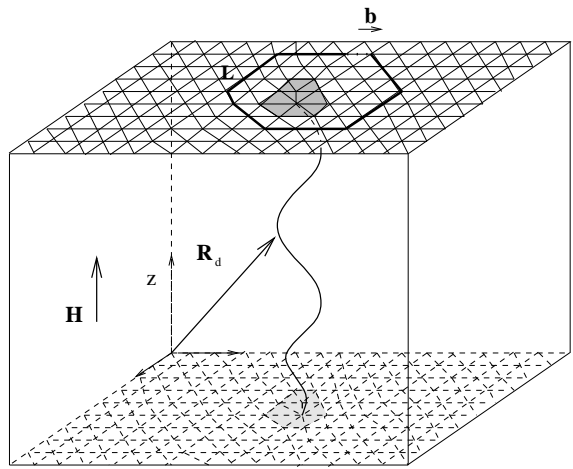} 
} 
\vspace*{0.2truecm}

\caption{Schematic plot of a TDL in a superconductor film.\\ 
The Burger's vector $\mbox{\boldmath $b$}$ lies in the plane
perpendicular to the\\magnetic field $\mbox{\boldmath $H$}$.}
\label{disloc} 
\vspace*{-0.3truecm}
\end{figure}
By comparing the experimental data with our results for mean square fluctuations 
of
a TDL, it is in principle possible to  estimate the elastic moduli of the vortex 
lattice.  
Hence, this analysis is complementary to that of the hydrodynamic model of 
a liquid of flux lines, used so far to quantify these 
coefficients\cite{Marchetti93}. 
On the other hand, the presence of even a single fluctuating TDL considerably
modifies the density correlation functions measured in the decoration 
experiments. The contribution of the fluctuating TDL to the long-wavelength 
structure factor is also {\em anisotropic} and involves the shear modulus, 
making it a good candidate for the determination of this coefficient.

In the usual experimental set-up, the magnetic field $\mbox{\boldmath $H$}$ 
is oriented along to the $z$ axis, perpendicular to the $CuO$-planes of the 
superconductor. The displacements of the flux lines from a perfect triangular
lattice at a point $(\mbox{\boldmath $r$},z)$, are described in the continuum 
elastic limit, by a {\em two-dimensional} vector field 
$\mbox{\boldmath $u$}(\mbox{\boldmath $r$},z)$.  
The corresponding elastic free-energy cost is 
\begin{equation}\label{2} 
{\cal H}=\int\frac{d^3\mbox{\boldmath $r$}}{2}
\left[c_{66}(\nabla\mbox{\boldmath $u$})^2+(c_{11}-c_{66})
(\nabla\cdot\mbox{\boldmath $u$})^2+c_{44}(\partial_z\mbox{\boldmath
$u$})^2\right],
\end{equation}
\noindent where $\nabla=(\partial_x\hat{x}+\partial_y\hat{y})$; and $c_{11}$, 
$c_{44}$, 
and $c_{66}$, are the compression, tilt, and shear elastic moduli, 
respectively.  
Due to the small magnetic fields involved in the experiments, non-local 
elasticity 
effects\cite{Blatter} are expected to be weak, and will be neglected for 
simplicity. 
In addition, at the temperatures corresponding to the freezing of decoration 
patterns, disorder-induced effects should be small, and will also be ignored. 

To describe a dislocation line, it is necessary to specify its position within
the material, and to indicate its character (edge or screw) at each point. 
The latter is indicated by the Burger's vector $\mbox{\boldmath $b$}$,
which in the continuum limit is defined by $\oint_L d\mbox{\boldmath 
$u$}=-\mbox{\boldmath $b$}$,
with $L$ a closed circuit around the dislocation \cite{Landau70}.
For the TDLs in our problem, the Burger's vectors lie in the $xy$-plane, and
the line conformations are generally described by the position vectors 
$\mbox{\boldmath $R$}_d(z)=\mbox{\boldmath $R$}(z)+z\hat{z}$
(see Fig. \ref{disloc}).
Unlike a vortex line, the wanderings of a TDL are highly {\em anisotropic}:
In an infinite system with a conserved number of flux lines, fluctuations
of the TDL are confined to the {\em glide plane} containing the Burger's 
vector and the magnetic field. The hopping of the TDL perpendicular to
its Burger's vector ({\em climb}) involves the creation of vacancy and 
interstitial lines, as well as the potential crossing of flux 
lines\cite{Labusch,Nabarro80}. These defects are very costly, 
making TDL climb unlikely, except, for instance, in the so-called 
{\em supersolid} phase, in which interstitials 
and vacancies are expected to proliferate \cite{Frey}.
Nevertheless, for a sample of finite extent, 
introduction and removal of flux lines from the edges may enable such motion.
This seems to be the case in some of the decoration experiments where the
number of flux lines is not the same on the two sides\cite{Yao94}.
In order to be completely general, at this stage we allow for the possibility
of transverse fluctuations in $\mbox{\boldmath $R$}(z)$, bearing in mind
that they may be absent due to the constraints.

We decompose the displacement field $\mbox{\boldmath $u$}$ into two parts: 
$\mbox{\boldmath $u$}^s\left(\mbox{\boldmath $r$}-\mbox{\boldmath 
$R$}(z),z\right)$, 
which represents the singular displacement due to a TDL passing through points 
$\mbox{\boldmath $R$}_d(z)$ in independent two-dimensional planes; 
and $\mbox{\boldmath $u$}^r$, a regular field due to the couplings between the 
planes. By construction, the former is the solution for a two-dimensional 
problem with 
the circulation constraint \cite{Landau70}, while the latter 
minimizes the 
elastic energy in Eq.(\ref{2}), and is consequently the solution to
\begin{displaymath} 
c_{66}\nabla^{2}\mbox{\boldmath $u$}^r+
(c_{11}-c_{66})\nabla \nabla\cdot\mbox{\boldmath $u$}^r + c_{44}
\partial^2_z\mbox{\boldmath $u$}^r = - c_{44} 
\partial^2_z\mbox{\boldmath $u$}^s.
\end{displaymath}
After Fourier transforming this equation and substituting its formal solution  
in Eq.(\ref{2}), the energy cost of a 
fluctuating 
TDL is obtained as ${\cal H}^d= {\cal H}^{str} + \Delta {\cal H}$, where 
${\cal H}^{str}$ the standard energy of a straight dislocation line. 
Keeping only the lowest order terms in the small deviations $\mbox{\boldmath 
$R$}(z)$
of the TDL axis from the straight line, gives the extra energy cost of 
distortions as
\begin{equation}\label{5}
\Delta {\cal H}=\int \frac{dk_z}{2\pi}
\left[A_{\perp}(k_z)|\mbox{\boldmath $R$}_{\perp}(k_z)|^2 +
A_{\parallel}(k_z)|\mbox{\boldmath $R$}_{\parallel}(k_z)|^2\right].
\end{equation}
Here, $\mbox{\boldmath $R$}(k_z)$ is the Fourier transform of 
$\mbox{\boldmath $R$}(z)$;  $\mbox{\boldmath $R$}_{\perp}$ and $\mbox{\boldmath
$R$}_{\parallel}$ stand for its components perpendicular and parallel to 
the Burger's vector, respectively; and

\end{multicols}
\begin{figure} 
\epsfxsize=17.5truecm
\vspace*{-0.9truecm}  
\leftline{
\epsfbox{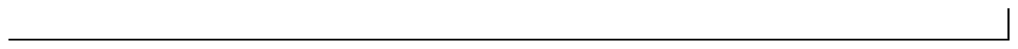} 
} 
\vspace*{-0.2truecm}
\end{figure}
\begin{eqnarray}\label{5a} 
A_{\perp}(k_z) &=&\frac{b^2c_{44}}{16\pi}k_z^2
\left[\ln \left(1+\frac{c_{66}\Lambda^2}{c_{44}k_z^2}\right)
+\left(2-4\frac{c_{66}}{c_{11}}+3\left(\frac{c_{66}}{c_{11}}\right)^{2}\right)
\ln \left(1+\frac{c_{11}\Lambda^2}{c_{44}k_z^2}\right)\right],\\ \label{5b}
A_{\parallel}(k_z) &=& \frac{b^2c_{44}}{16\pi}k_z^2
\left[\ln\left(1+\frac{c_{66}\Lambda^2}{c_{44}k_z^2}\right)+
\left(\frac{c_{66}}{c_{11}}\right)^{2}\ln
\left(1+\frac{c_{11}\Lambda^2}{c_{44}k_z^2}\right)\right].  
\end{eqnarray}
\begin{figure} 
\epsfxsize=21.2truecm 
\vspace*{-0.6truecm} 
\leftline{
\epsfbox{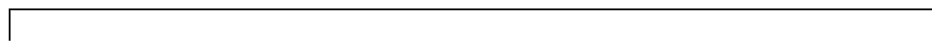}
}
\vspace*{-0.6truecm}  
\end{figure}
\begin{multicols}{2}
\noindent The above expressions are obtained after integrating over 
$\mbox{\boldmath $q$}$, 
with a long-wavevector cutoff $\Lambda$ at distances of  the order of the 
flux-line lattice spacing, below which the continuum treatment is not valid. 
If we also take into account a short-wavevector cutoff $\Lambda^*$ due to finite
sample area, the dependence of the kernels on $k_z$ has different forms. For 
values of $k_z \ll\Lambda^* \sqrt{\min(c_{66},c_{11})/c_{44}}$, the logarithms 
in 
Eqs.(\ref{5a}-\ref{5b}) reduce to the constant value $2\ln(\Lambda/\Lambda^*)$, 
and both kernels are simply proportional to $k_z^2$. In the opposite limit, if 
$k_z\gg\Lambda\sqrt{\max(c_{66},c_{11})/c_{44}}$ all the logarithms can be 
approximated by the first term of their Taylor expansion, and $A_{\perp}$ and 
$A_{\parallel}$ are independent of $k_z$. In between these extremes,
the form of the kernels is globally represented through Eqs.(\ref{5a}-\ref{5b}). 
In practice, the smallest wavevector $k_z$ that can be probed experimentally 
is limited by the finite thickness of the sample, and is ultimately constrained 
(by the measured values of $c_{11}$, $c_{66}$ and $c_{44}$)  to the last two 
regimes.
From Eqs.(\ref{5}-\ref{5b}), we conclude that the TDL behaves as an elastic 
string with a dispersive line tension ($\epsilon_d\propto \ln k_z$), indicating 
a non-local elastic energy.  (A single flux line shows a similar dispersive 
behavior, as pointed out by Brandt\cite{Brandt76}.)

Equilibrium thermal fluctuations of a TDL are calculated from 
Eq.(\ref{5}), {\em assuming} that one can associate the 
Boltzmann probability $e^{-\Delta {\cal H}/k_BT}$ to this metastable state. 
After averaging over all possible configurations of 
$\mbox{\boldmath $R$}(k_z)$, the mean square displacements are obtained as
\begin{equation}\label{6} 
\langle|R_{\perp}(k_z)|^2\rangle=\frac{k_BTL}{A_{\perp}(k_z)}, \mbox{ and } 
\langle|R_{\parallel}(k_z)|^2\rangle= \frac{k_BTL}{A_{\parallel}(k_z)}, 
\end{equation}
respectively, where $L$ is the thickness of the film. In terms of  the function
\begin{equation}\label{8b} 
\langle |R(k_z)|^2\rangle_{c_{11}}\equiv\frac{8\pi k_BTL}{b^2 c_{44} k_z^2}
\ln^{-1}\left(1+\frac{c_{11}\Lambda^2}{c_{44}k_z^2}\right),  
\end{equation}
these quantities satisfy the simple relation
\begin{equation}\label{8} 
\langle |R_{\perp}|^2\rangle^{-1}-\langle
|R_{\parallel}|^2\rangle^{-1}= \langle |R|^2\rangle_{c_{11}}^{-1}
\left(1-\frac{c_{66}}{c_{11}}\right)^2.  
\end{equation}
Thus, even if the TDL is allowed to meander without constraints, its 
fluctuations
are {\em anisotropic}, as $A_{\perp}(k_z)= A_{\parallel}(k_z)$ only for 
$c_{11}=c_{66}$. 
In HTSC materials,  $c_{66}\ll c_{11}$, so that 
$A_{\perp}(k_z)>A_{\parallel}(k_z)$, 
limiting fluctuations largely to the glide plane.

In real space, the width of the TDL depends on quantities such as
$\langle |R(L)|^2\rangle_{c_{11}} \equiv 1/L \int dk_z/2\pi \ 
\langle |R(k_z)|^2\rangle_{c_{11}}$.
Its dependence on the film thickness $L$ follows from Eq.(\ref{8b}), as
\begin{equation}\label{8d} 
\langle |R(L)|^2\rangle_{c_{11}}\sim \frac{k_BT}{c_{11}} 
\left\{ \begin{array}{ll} 1/d_o & \mbox{\rm if } L\ll l_{1},\\ 
{L}/\left[{2\ l_1^2} \ln\left(\frac{L}{l_1}\right)\right] & \mbox{\rm if }
L\gg l_{1} ,\end{array} \right.\nonumber  
\end{equation}
where we have defined $l_{1}\equiv b\sqrt{c_{44}/c_{11}}$, and $d_o$
is a short-distance cutoff along the $z$-axis.  For length-scales below $d_o$,
the layered nature of the material is important.

\begin{figure} 
\epsfxsize=8truecm 
\centerline{
\epsfbox{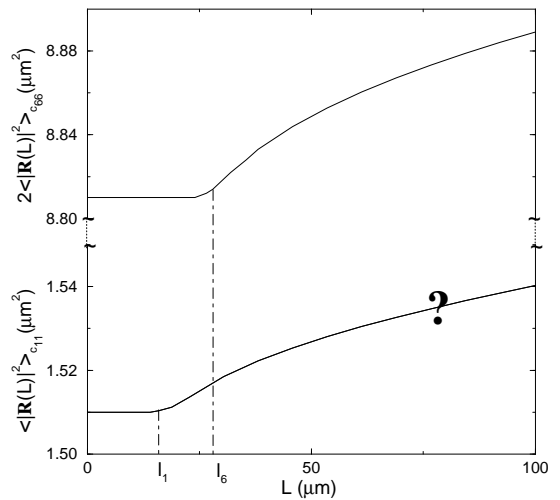} 
} 

\caption{Mean-square displacements as a function of the\\
thickness $L$, both measured in $\mu m$, for a BSCCO sample\\ 
decorated in a field of $24 G$ ($a_o=b=1\mu m$).}
\label{real.ps} 
\vspace*{-0.3truecm}
\end{figure}

In order to estimate typical fluctuations for the TDL, we assume that
$c_{66}\ll c_{11}$, and approximate Eq.(\ref{5b}) by its leading behavior.
In this limit,  $\langle |R_{\parallel}|^2\rangle \simeq 2 \langle 
|R|^2\rangle_{c_{66}}$,
with $\langle |R|^2\rangle_{c_{66}}$ defined as in Eq.(\ref{8b}), after 
replacing 
$c_{11}$ with $c_{66}$.  A behavior similar to Eq.(\ref{8d}) is obtained for 
$\langle |R|^2\rangle_{c_{66}}$, with a corresponding crossover length 
$l_{6}\equiv b \sqrt{c_{44}/c_{66}}$.  Thus, longitudinal fluctuations of the 
TDL are approximately constant for samples thinner than $l_{6}$, and grow as 
$L/\ln L$ for thicker samples.  If allowed, transverse fluctuations follow from
Eq.(\ref{8}), whose leading behavior
for small $c_{66}$ gives, $\langle |R_{\perp}|^2\rangle\sim \langle
|R|^2\rangle_{c_{11}}$.  In Fig.\ref{real.ps} we have plotted $\langle
|R|^2\rangle_{c_{11}}$ and $2\langle |R|^2\rangle_{c_{66}}$ as a function of
the thickness $L$.  Both quantities are very sensitive to the elastic 
coefficients. We have considered the values of $c_{11}=2.8\times 10^{-2} G^2$, 
$c_{44}=8.1 G^2$, and $c_{66}=9.6\times 10^{-3}G^2$ reported in
Ref.\cite{Yoon95} for a sample decorated in a field of $24 G$, 
which shows a single dislocation.  The Burger's
vector is equal to the average lattice spacing $a_o=1\mu m$, and the
short-distance cutoff in the plane is taken to be of the same order of 
magnitude.  The crossover lengths introduced turn out to be 
$l_{1}\sim 17\mu m$, and $l_{6}\sim 29\mu m$, so that the 
experimental sample thickness ($L\sim 20 \mu  m$)
approximately falls into the constant regime in Fig.\ref{real.ps}.  
From the top curve in Fig.\ref{real.ps}, we estimate 
$\langle |R_{\parallel}|^2\rangle^{1/2}\sim 3\mu m$ for $T\sim 80 K$, with an 
uncertainty factor of about $\sqrt{10}$ due to, for instance, the uncertain
values of temperature, and both the in-plane and perpendicular cutoffs. 
If unconstrained, transverse fluctuations of the TDL are smaller, and given by 
$\langle |R_{\perp}|^2\rangle^{1/2}\sim 1\mu m$, in the same regime.
The question mark in Fig.\ref{real.ps} is a reminder that once these 
fluctuations 
exceed a lattice spacing, proper care must be taken to  account for constraints,
and their violation by defects or surface effects.

As discussed in Ref.\cite{Yoon95}, the values of $c_{11}$ and $c_{66}$ measured 
in the experiments are about three orders of magnitude smaller than the 
theoretical predictions from Ginzburg-Landau theory. If we use the latter 
values in our computations, the crossover lengths become much shorter and 
TDL fluctuations are reduced by three orders of magnitude! 
Due to this sensitivity, analysis of transverse and longitudinal 
fluctuations of TDLs in two-sided decoration experiments should provide
a complementary method for determining the elastic moduli.
Unfortunately, the films studied so far are in the short distance regime where 
details of the cutoff play a significant role. Experiments on thicker films are 
needed to probe the true continuum limit.

TDLs also produce anisotropies in the flux line density $n(\mbox{\boldmath 
$r$},z)$, 
and the corresponding diffraction patterns. Neutron scattering studies can in 
principle
resolve the full three dimensional structure factor $S(\mbox{\boldmath 
$q$},k_z)=
\left\langle |n(\mbox{\boldmath $q$},k_z)|^2 \right\rangle$, although only a few 
experiments are currently available for different HTSC 
materials\cite{Cubbit93,Yaron94}. 
Two-sided decoration experiments also provide a quantitative characterization of 
the
{\it two-dimensional} structure factors calculated from each surface, as well as 
the correlations between the two sides of the sample. 

The diffraction pattern from a vortex solid has Bragg peaks at the reciprocal 
lattice positions.
Unbound dislocations modify the translational correlations; a finite 
concentration
of dislocation loops can drive the long wavelength shear modulus to zero, while 
maintaining the long-range orientational order \cite{Marchetti90}. The resulting 
hexatic phase has diffraction rings with a 6-fold modulation, which disappears 
in the liquid phase.
In all phases, the diffuse scattering close to  $\mbox{\boldmath $q$}=0$ is 
dominated by the long wavelength density fluctuations, which are adequately 
described by $n=\nabla\cdot\mbox{\boldmath $u$}$, leading to 
\begin{equation}\label{11} 
S(\mbox{\boldmath $q$},k_z)\sim
\langle|\mbox{\boldmath $q$}\cdot\mbox{\boldmath $u$}(\mbox{\boldmath
$q$},k_z)|^2\rangle.  
\end{equation}
The contribution of equilibrium density fluctuations (from longitudinal phonons) 
to Eq.(\ref{11}) has the form\cite{Marchetti93}
\begin{equation}\label{phononS} 
S^o(\mbox{\boldmath $q$},k_z)=\frac{k_BTLAq^2}{c_{11}q^2+c_{44}k_z^2},  
\end{equation}
where $A$ is the sample area. This contribution is clearly isotropic, and 
independent of the shear modulus in the solid phase.
(The anisotropies of the solid and hexatic phases are manifested at higher 
orders in $\mbox{\boldmath $q$}$.)
For a sample of finite thickness, the phonon contribution in rather general 
situations including surface and disorder effects, was obtained in 
Ref.\cite{Marchetti93}, as 
$S^o(\mbox{\boldmath $q$},L)=S_{2D}^o(\mbox{\boldmath $q$})
R(\mbox{\boldmath $q$},L)$, where $S_{2D}^o(\mbox{\boldmath $q$})$ is the 
$2D$ structure factor of each surface, while $R(\mbox{\boldmath $q$},L)$ 
measures the correlations between patterns at the two sides of the film.

The above results were used in Ref.\cite{Yoon95} to determine the elastic
moduli $c_{11}$ and $c_{44}$ of the vortex array, at different magnetic fields. 
However, the decoration images used for this purpose have the appearance of
a solid structure with a finite number of topological defects. We shall 
demonstrate here that the presence of a single trapped TDL modifies the 
isotropic behavior in Eq.(\ref{phononS}).
We henceforth decompose the displacement field
$\mbox{\boldmath $u$}(\mbox{\boldmath $r$})$ into a  {\em regular} phonon part 
$\mbox{\boldmath $u$}^o$, and a contribution 
$\mbox{\boldmath $u$}^d=\mbox{\boldmath $u$}^s+\mbox{\boldmath $u$}^r$ from the 
meandering TDL described by $\mbox{\boldmath $R$}(z)$. 
The overall elastic energy also decomposes into independent contributions
${\cal H}[\mbox{\boldmath $u$}^o,\mbox{\boldmath $R$}]=
{\cal H}^o[\mbox{\boldmath $u$}^o]+{\cal H}^d[\mbox{\boldmath $R$}]$. 
To calculate the average of any quantity, we integrate over smoothly varying
displacements $\mbox{\boldmath $u$}^o$, and over distinct configurations of the
dislocation line $\mbox{\boldmath $R$}(z)$.  Thus the structure factor in 
Eq.(\ref{11}) becomes a sum of {\em phonon} and {\em dislocation} parts.  

The results of Eq.(\ref{6}) can be used to calculate the contribution from a 
fluctuating TDL, which has the form
\end{multicols}
\begin{figure} 
\epsfxsize=17.5truecm
\vspace*{-0.9truecm}  
\leftline{
\epsfbox{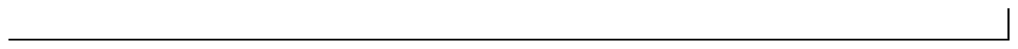} 
}
\vspace*{-0.2truecm} 
\end{figure}
\begin{equation}\label{14} 
S^d(\mbox{\boldmath $q$},k_z)=
\frac{8\pi L b^2 c_{66}^2 q_{\perp}^2}{c_{11}^2q^4}\delta(k_z)+
\frac{k_BTL b^2}{(c_{11}q^2+c_{44}k_z^2)^2}
\left(\frac{4c_{66}^2q_{\parallel}^2q_{\perp}^2}{A_{\parallel}(k_z)}+
\frac{(c_{44}k_z^2+2c_{66}q_{\perp}^2)^2}{A_{\perp}(k_z)}\right), 
\end{equation}
\begin{figure} 
\epsfxsize=21.2truecm 
\vspace*{-0.6truecm} 
\leftline{
\epsfbox{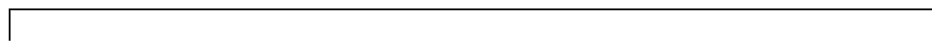} 
}
\vspace*{-0.6truecm}  
\end{figure}
\begin{multicols}{2}

\noindent with $q_{\parallel}\equiv \mbox{\boldmath $q$}\cdot \mbox{\boldmath 
$b$}/b$, and $q_\perp$ the component perpendicular to the Burger's vector. The 
first term on the r.h.s. of Eq.(\ref{14}) corresponds to the straight TDL, 
and vanishes in the liquid state with $c_{66}=0$. 
The next two terms on the r.h.s. result from the longitudinal and transverse 
fluctuations of the TDL respectively. 
(The latter is absent if the TDL is constrained to its glide plane.) 
The {\em dislocation} part is clearly anisotropic, and the anisotropy involves 
the shear modulus $c_{66}$. Thus after inverting the $k_z$ transform in 
Eq.(\ref{14}), the TDL contribution to the structure factors calculated from the 
two-sided decoration experiments can also be exploited to obtain information
about the elastic moduli.

In conclusion, we have calculated the energy cost of meanderings of a
TDL in the flux lattice of a HTSC film such as  BSCCO. Flux decoration
experiments indicate that such metastable TDLs are indeed frequently
trapped in thin films in the process of field cooling. We have estimated the
thermal fluctuations of a TDL in crossing the sample, as well as its 
contribution to the structure factor. Both effects can in principle be used to
estimate the elastic moduli of the vortex solid. However, there are strong
interactions between such defects, which need to be considered when a
finite number of TDLs of different Burger's vectors are present. 
The generalization of the approach
presented here to more than one TDL may provide a better description of the
experimental situation. From the experimental perspective, it should be 
possible to find samples with a single trapped TDL, providing a direct test of
the theory.  Other realizations of TDLs can be found in grown films\cite{TDLs}, 
and may also occur in smectic liquid crystals. 
It would be interesting to elucidate the similarities and distinctions 
between the defects in these systems.

MCM acknowledges financial support from the Direcci\'o General de Recerca
(Generalitat de Catalunya). MK is supported by the NSF Grant No. DMR-93-03667. 
We are grateful to D.R. Nelson for emphasizing to us the constraints on 
transverse motions of TDLs.
We have benefited from conversations with M.V. Marchevsky, and also thank Z. 
Yao for providing us with the raw decoration images in Refs.\cite{Yao94,Yoon95}.

\end{multicols}

\vspace*{-0.5truecm}

\begin{references}

\vspace*{-1.5truecm}

\bibitem{Yao94} Z.  Yao {\em et al.}, Nature {\bf 371}, 777 (1994).

\bibitem{Yoon95} S.  Yoon {\em et al.}, Science {\bf 270}, 270 (1995).

\bibitem{Blatter} G.  Blatter {\em et al.}, Rev.  Mod.  Phys.  {\bf 66}, 1125
(1994).

\bibitem{Grier} D.  Grier {\em et al.}, Phys.  Rev.  Lett.  {\bf 66}, 2270
(1991).

\bibitem{Marchevsky} M.V. Marchevsky, PhD Thesis, Leiden University (1997).

\bibitem{TDLs} E.A. Fitzgerald, Mater. Sci. Rep. {\bf 7}, 87 (1991).

\bibitem{Labusch} R.  Labusch, Phys.  Lett.  {\bf 22}, 9 (1966).

\bibitem{Nabarro80} F.R.N.  Nabarro and A.T.  Quintanilha, in {\it Dislocations
in Solids}, edited by F.R.N.  Nabarro (North-Holland, Amsterdam, 1980), Vol.  5.

\bibitem{Frey} E. Frey, D.R. Nelson, and D.S. Fisher, Phys. Rev. B {\bf 49}, 
9723 (1994).

\bibitem{Marchetti93} M.C.  Marchetti and D.R.  Nelson, Phys. Rev. B {\bf 47},
12214 (1993); Phys. Rev. B {\bf 52}, 7720 (1995).

\bibitem{Landau70} L.D.  Landau and E.M.  Lifschitz, {\it Theory of Elasticity}
(Pergamon, New York, 1970).

\bibitem{Brandt76} E.H.  Brandt, J.  Low Temp.  Phys.  {\bf 26}, 735 (1977).

\bibitem{Marchetti90} M.C.  Marchetti and D.R.  Nelson, Phys.  Rev.  B {\bf 41},
1910 (1990).

\bibitem{Cubbit93} R.  Cubbit {\em et al.}, Nature {\bf 365}, 407 (1993).

\bibitem{Yaron94} U.  Yaron {\em et al.}, Phys.  Rev.  Lett.  {\bf 73}, 2748
(1994).

\end{references}
\end{document}